\documentstyle[11pt,paspconf]{article}
\begin{document}

\title{Ages and Chemical Abundances in Dwarf Spheroidal Galaxies}
\author{Tammy Smecker-Hane\altaffilmark{1}}
\affil{Dept. of Physics \& Astronomy, 4129 Reines Hall,
      University of California, Irvine, CA 92697-4575}
\author{Andrew McWilliam}
\affil{Observatories of the Carnegie Institute of Washington,
       813 Santa Barbara St., Pasadena, CA 91101-1292}

\begin{abstract}
The dwarf spheroidal galaxies (dSphs) in the Local Group are
excellent systems on which we can test theories of galaxy
formation and evolution. Color-magnitude diagrams (CMDs)
containing many thousands of stars from the asymptotic giant
branch to well below the oldest main-sequence turnoff are being
used to infer their star-formation histories, and surprisingly
complex evolutionary histories have been deduced.  Spectroscopy of
individual red giant stars in the dSphs is being used to determine
the distribution of chemical abundances in them. By combining
photometry and spectroscopy, we can overcome the age-metallicity
degeneracy inherent in CMDs and determine the evolution of dSphs
with unprecedented accuracy. We report on recent progress and
discuss a new and exciting avenue of research, high-dispersion
spectroscopy that yields abundances for numerous chemical
elements.  The later allows us to estimate the enrichment from
both Type~Ia and Type~II supernovae (SNe) and places new limits on
how much of the Galaxy could have been accreted in the form of
dSph-sized fragments and when such mergers could have taken place.
\end{abstract}


\altaffiltext{1}{Visiting Astronomer at the W.\ M.\ Keck
Observatory and the Cerro Tololo Inter-American Observatory. KO is
operated as a scientific partnership among the California Inst. of
Technology, the Univ. of California and NASA, and was made
possible by the generous financial support of the W.\ M.\ Keck
Foundation. CTIO is operated by AURA, Inc.\ under cooperative
agreement with the NSF. Funding for a large part of the research
discussed herein was provided by NSF grant AST-9619460 to TSH.}

\section{Introduction}

We begin with a review of the typical physical characteristics of
a dSph galaxy. DSphs have low total luminosities ($-9 \geq M_V
\geq -14$) and small physical sizes (core radii $\approx$ 200 to
600 pc). They are stellar systems with little or no detectable
interstellar medium. Upper limits on the mass of HI inside the
optical radius are typically $<10^4$ M$_\odot$ (Young 1999, and
references therein). However, Carignan et al.\ (1998) recently
discovered at least $3 \times 10^4$ M$_\odot$ of HI lying just
outside the optical radius of the Sculptor dSph. Additional
searches should be made for gas surrounding the Fornax and Leo I
dSphs, which have very young stars, $\sim0.1$ Gyr and 1 Gyr,
respectively, in their CMDs (Stetson, et al. 1998, Gallart, et al.
1999). The dSphs in the Local Group are nearby and diffuse
(central surface densities $\approx$ 0.1 to 1 M$_\odot$/pc$^2$),
which makes them very amenable to photometric studies.

Low total masses ($10^7$ to $10^9$ M$_\odot$) have been inferred
for dSphs from their small central stellar velocity dispersions (7
to 14 km/s). (See Mateo 1998 for a recent review.) The masses are
usually inferred assuming mass follows light. However, if dark
matter halos are more extended than the luminous galaxies, as is
probably the case, then these are only lower limits to the total
mass. Whether or not the mass inside the optical radius is
dominated by stars or dark matter is a function of galaxy
luminosity (Mateo 1998). The derived mass-to-light ratios vary
from $M/L_V = 5$ (in solar units) for the Fornax dSph ($M_V =
-14$) to $M/L_V = 200$ for the Draco dSph ($M_V = -9$). The former
being only slightly larger than the $M/L_V = 4$ expected for an
old stellar population, and the later indicating a mass dominated
by dark matter. The physical cause of the clear trend in $M/L_V$
versus $L_V$ is an open question. Differences in total mass or
concentration of the dark matter halos could make some dSphs more
susceptible to loosing gas in supernova-driven galactic winds and
thus less efficient at converting their gas into stars.

\section{Star-Formation Histories}

Great strides have been made recently in quantifying the ages of
stars in dSphs through CMD analysis because of the increased
availability of wide-field CCD cameras on 4-meter class telescopes
and the high angular resolution of WFPC2 on the Hubble Space
Telescope. Two examples are shown in Figure 1. Much to our initial
surprise, we found most of the dSphs in the Local Group have had
complex star-formation histories! (See Smecker-Hane 1997 for a
recent review.) Nearly all of the dSphs began forming stars $\sim
14$ Gyr ago during the epoch of globular cluster formation in our
Galaxy. Some dSphs lost their gas and stopped forming stars
relatively quickly (Ursa Minor: Hernandez, et al.\ 1999; Draco:
Grillmair, et al.\ 1998), some dSphs continued forming stars for
many Gyr (Leo II: Mighell \& Rich 1996, Carina: Smecker-Hane, et
al. 1999a, Hurley-Keller, et al.\ 1998), other dSphs continued to
vigorously form stars until only 1 or 2 Gyr ago (Fornax: Stetson,
et al.\ 1998; Leo I: Gallart, et al.\ 1999). Carina is a
particularly intriguing dSph because it stopped forming stars for
a few Gyr ($\sim 7$ to 10 Gyr ago) after which star formation
resumed again for another few Gyr ($\sim 7$ to 4 Gyr ago). Thus
most of the dSphs formed stars over many dynamical timescales (few
$\times$ 0.1 Gyr) despite their low velocity dispersions ($\sim
10$ km/s). They were able to retain and recycle gas even though
thousands of SNe exploded in them. A key observation may be that
dSphs form stars slowly. It may be that massive stars, through
photoionization and stellar winds, expand networks of bubbles and
tunnels through the interstellar medium (ISM) so the hot ejecta
from SNe quickly escapes the galaxy without imparting much kinetic
energy to the ISM (e.g.\ Lin \& Murray 1993). Thus dSphs may be
teaching us that star formation is self-regulated even in the
lowest mass galaxies.

\begin{figure}
\plotfiddle{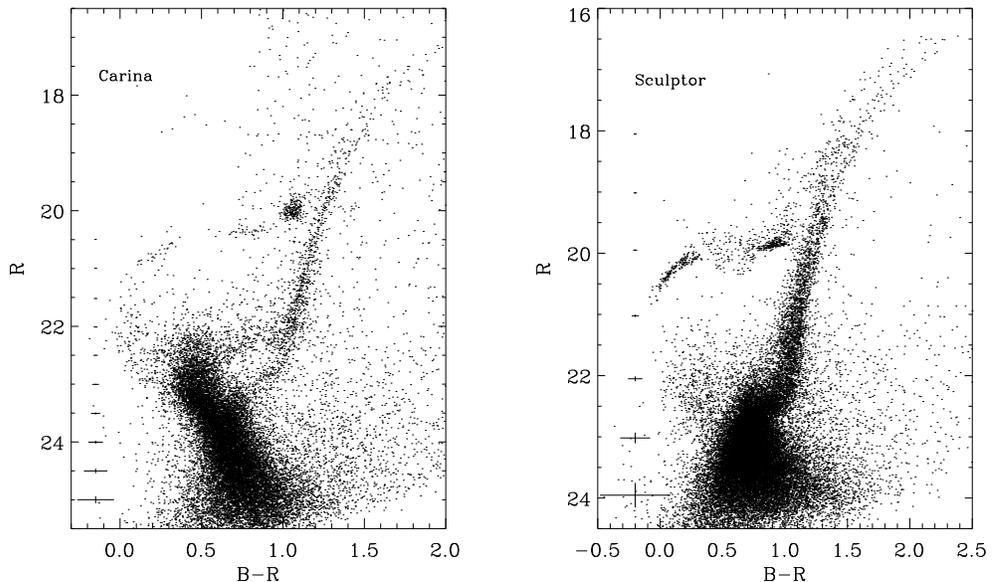}{7cm}{0.}{70.}{70.}{-210}{-25}
\vspace{0.6cm} \caption{CMDs for the Carina dSph (left;
Smecker-Hane, et al.\ 1999a) and the Sculptor dSph (right;
Smecker-Hane, et al.\ 1999b) obtained with the CTIO
4.0-meter telescope.}
\end{figure}

\section{Chemical Abundances}

Measuring the chemical abundances of the stellar populations in
dSphs through spectroscopy of individual red giant stars will be
crucial for constraining the inflow and outflow of gas from these
galaxies. (For example, was star formation in Carina renewed by
infall of fresh gas, or did gas remain gravitationally bound the
dSph but was temporarily suspended from cooling and forming stars
by something such as the intra-galactic UV background?)
Independently determining the metallicities of the stars via
spectroscopy also gives us the ability to overcome the
age-metallicity degeneracy inherent in CMDs and enables us to
infer unique and accurate solutions for star-formation histories.

Studies of the chemical abundances of stars in dSphs have been
made, but the next few years will be particularly fruitful because
of the wide availability of multi-fiber spectrographs. The method
most commonly used to measure metallicities of red giant stars in
dSphs is the reduced equivalent width of the calcium infrared
triplet (see Rutledge, et al.\ 1997, and references therein). This
method is preferred because relative metallicities accurate to 0.1
dex can be obtained for spectra with moderate resolution ($R
\approx 3500$) and moderate signal-to-noise ratio
(S/N$\approx30$). DSphs have low mean metallicities, [Fe/H]
$\approx -2.0$ dex, although each dSph has a significant internal
metallicity dispersion approximately described by a Gaussian with
$\sigma \simeq 0.25$ dex with the full range in metallicities
spanning $\sim 1.0$ dex (Draco: Lehnert, et al.\ 1992, Sextans:
Suntzeff, et al.\ 1993, Carina: Smecker-Hane, et al.\ 1999c). This
implies recycling of SNe ejecta and an extended period of star
formation ($> 0.1$ Gyr) for even the Draco dSph, which contains
primarily $\sim 14$ Gyr old stars.

\section{Chemical Abundance Ratios}

The complex star-formation histories inferred for most dSphs
should leave obvious signatures in the chemical element ratios of
their stars. Consider a delta-function burst of star formation.
Type~II SNe will explode after $<$ few $\times$ $10^7$ yr, and
enrich the ISM with a high fraction of $\alpha$ elements (e.g.\,
O, Ca, Mg, Si, Ti) relative to Fe--peak elements (e.g.\ Fe, Cr,
Ni). For example, the IMF--averaged yield from Type~II SNe has
[Ca/Fe] $ =+0.41$ (Timmes, et al.\ 1995). In contrast, Type~Ia SNe
begin exploding approximately 0.1 Gyr after the stars form and
continue for many Gyr. However, their ejecta is dominated by
Fe--peak elements. The yield of Type~Ia SNe has [Ca/Fe] $=-0.31$
(Thielemann, et al.\ 1986). This has interesting consequences for
constraining how much of the Galaxy could be built through
accretion of dwarf-galaxy sized fragments, and when such mergers
could have occurred. For example, typical Galactic halo stars have
[$\alpha$/Fe] approximately equal to the Type~II SNe yields (see
McWilliam 1997, and references therein). Thus the Galactic halo
has been inferred to have formed quickly because only ejecta from
short-lived Type~II SNe, and not from long-lived Type~Ia SNe, were
incorporated into halo stars.

The Sagittarius (Sgr) dSph was recently discovered to be merging
with the Galaxy (Ibata, et al.\ 1997), although how fast this
merger is proceeding is debated. At a distance of a only 24 kpc,
Sgr dSph stars are amenable to high-dispersion spectroscopy and
abundances of numerous chemical species can be measured. We
obtained echelle spectra with HIRES (Vogt, et al.\ 1994) on the
Keck~I telescope for 14 stars in the Sgr dSph that span a wide
range of age and metallicity as inferred from their position the
CMD (Sarajedini \& Layden 1995). Our echelle spectra cover 5200 to
7600 \AA and have high resolution ($R = 43000$) and high
signal-to-noise ratios (40 to 80). The spectra were reduced with
the standard IRAF ECHELLE package, and equivalent widths for
approximately 270 absorption lines were measured with our GETJOB
program (McWilliam, et al.\ 1995). A sample spectrum is shown in
Figure 2.

\begin{figure}
\plotfiddle{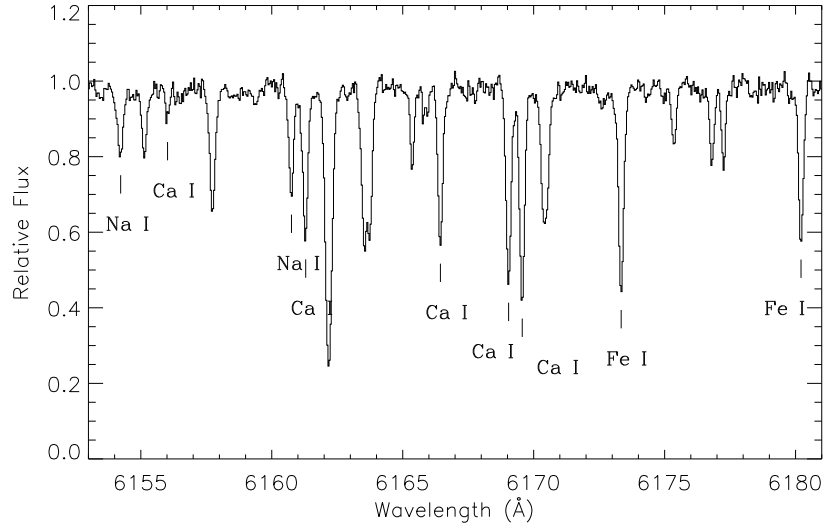}{6cm}{0.}{60.}{60.}{-175}{-240}
\vspace{0.7cm} \caption{A section of the echelle spectrum of
Sagittarius dSph star \#1-73, whose abundances are [Fe/H] $=-1.41$
and [Ca/Fe]$=+0.21$.}
\end{figure}

To derive chemical abundances, we use model atmospheres from
Kurucz (1992) and a heavily modified version of MOOG (Sneden
1973). We began by adopting initial effective temperatures
($T_{eff}$) and gravities (log $g$) determined from VIJK-band
photometry. However the final stellar parameters are determined
self-consistently from the model atmosphere analysis. We set log
$g$ by requiring consistency between the abundances derived from
Fe I and Fe II lines, $T_{eff}$ by requiring abundances derived
from Fe I lines be independent of excitation potential, and the
microturbulence by requiring abundances derived from Fe I lines be
independent of equivalent width. (Differences between initial and
final $T_{eff}$ values ranged from 0 to 300 $^\circ$K.) The
preliminary abundance results for our first 11 stars are shown in
Table 1. The mean $1-\sigma$ measurement errors are typically 0.05
dex for [Fe/H] and 0.1 dex for [X/Fe]. We are investigating
potential non-LTE effects on the Ti abundances derived for the 2
most metal-poor stars. We also will derive abundances for
neutron-capture elements such as Eu, Ba and La, whose line
profiles are non-Gaussian because of hyper-fine splitting, through
spectral synthesis.

\begin{table}
\caption{Preliminary chemical abundances derived for Sagittarius
dSph stars (Smecker-Hane \& McWilliam 1999). The adopted solar
abundances are the meteoric values from Anders \& Grevesse
(1989).}
\begin{center} \scriptsize
\begin{tabular}{|c|c|ccccc|}
\hline STAR  & [Fe/H]& [O/Fe]&[Na/Fe]&[Al/Fe]&[Ca/Fe]&[Si/Fe]
\\ \hline
   1--87  & --1.41 &  +0.19 & --0.08 & --0.01 &  +0.31 &  +0.20
\\ 1--73  & --1.14 & --0.82 &  +0.23 &  +1.20 &  +0.21 &  +0.50
\\ 1--150 & --0.59 & --0.08 & --0.53 & --0.33 &  +0.06 &  +0.02
\\ 1--245 & --0.59 & --0.00 & --0.32 &  +0.13 &  +0.10 &  +0.10
\\ 2--38  & --0.57 & --0.07 & --0.56 & --0.27 &  +0.08 & --0.03
\\ 1--229 & --0.52 & --0.19 & --0.51 & --0.38 & --0.04 &  +0.07
\\ 1--242 & --0.40 & --0.14 & --0.31 & --0.30 & --0.14 &  +0.16
\\ 2--75  & --0.37 & --0.09 & --0.31 & --0.17 &  +0.08 & --0.02
\\ 1--95  & --0.26 &  +0.08 & --0.37 & --0.24 & --0.11 &  +0.00
\\ 1--267 & --0.06 & --0.20 & --0.51 & --0.41 & --0.03 &  +0.08
\\ 2--85  & --0.03 &  +0.00 & --0.34 & --0.33 &  +0.03 &  +0.05
\\ \hline
\end{tabular}

\vspace{2ex}

\begin{tabular}{|c|c|cccc|}
\hline STAR  & [Fe/H]& [TiI/Fe] & [TiII/Fe] & [Cr/Fe] & [Ni/Fe]
\\ \hline
   1--87  & --1.41 &  +0.14 &  +0.43 &  +0.05 & --0.09
\\ 1--73  & --1.14 & --0.09 &  +0.49 &  +0.03 &  +0.01
\\ 1--150 & --0.59 & --0.17 &  ---   & --0.20 & --0.09
\\ 1--245 & --0.59 & --0.07 &  +0.10 &   ---  & --0.17
\\ 2--38  & --0.57 & --0.01 & --0.01 & --0.14 & --0.14
\\ 1--229 & --0.52 & --0.09 &  +0.05 & --0.18 & --0.18
\\ 1--242 & --0.40 &  +0.02 &  +0.33 &   ---  & --0.09
\\ 2--75  & --0.37 &  +0.11 &  +0.01 & --0.16 & --0.06
\\ 1--95  & --0.26 &  +0.08 &  +0.15 &   ---  &   ---
\\ 1--267 & --0.06 &  +0.04 &  +0.04 & --0.05 &  +0.00
\\ 2--85  & --0.03 & --0.03 & --0.06 &   ---  & --0.14
\\ \hline
\end{tabular}
\end{center}
\end{table}

In a similar study, Brown, et al. (1999) obtained echelle spectra
at lower resolution ($R=24000$) with the CTIO 4-meter telescope
for 2 of the brightest red giants in the globular cluster M54, a
member of the Sgr dSph. Also, Shetrone, et al.\ (1998) obtained
lower signal-to-noise (S/N$=27$) spectra with HIRES on the Keck~I
telescope for 4 stars in the Draco dSph.

Figure 3 illustrates the variation of [Ca/Fe] with metallicity for
dSphs stars, and Figure 4 shows the same for Galactic stars.
(Other $\alpha$ elements show similar behaviour.) Surprisingly,
the trend of [$\alpha$/Fe] with [Fe/H] for the Sgr dSph more
resembles that of the Galactic disk than the Galactic halo! The
relatively smooth trend ending with stars having solar metallicity
and solar element ratios implies a long, continuous rather than
episodic, star-formation history. Seventy-five percent of the iron
in stars with [Ca/Fe] $= 0$ was synthesized in Type~Ia SNe. Thus
the star-formation rate and chemical evolution of the Sgr dSph was
complex. We suggested the solar metallicity red giants in the Sgr
dSph were very young, $\sim 1$ Gyr old, and Bellazzini, et al.\
(1999) have recently found the possible main-sequence counter
parts of these red giants in their CMD. Note that the current
metallicity of the Large Magellanic Cloud is significantly lower,
[Fe/H] $\approx -0.3$, than the most metal-rich stars in the Sgr
dSph even though the LMC is approximately 45 times more luminous
that the Sgr dSph! Could this mean a large fraction of Sgr dSph's
stars were stripped away recently, or does the Sgr dSph accrete
gas from the Galactic disk when it plunges through it? Dynamical
arguments seem to disfavor both ideas leaving us to hypothesize
that the core of the dark matter halo may be massive enough for
star formation to continue in the central regions while the
lower-density outer regions are slowly peeled away.

\begin{figure}
\plotfiddle{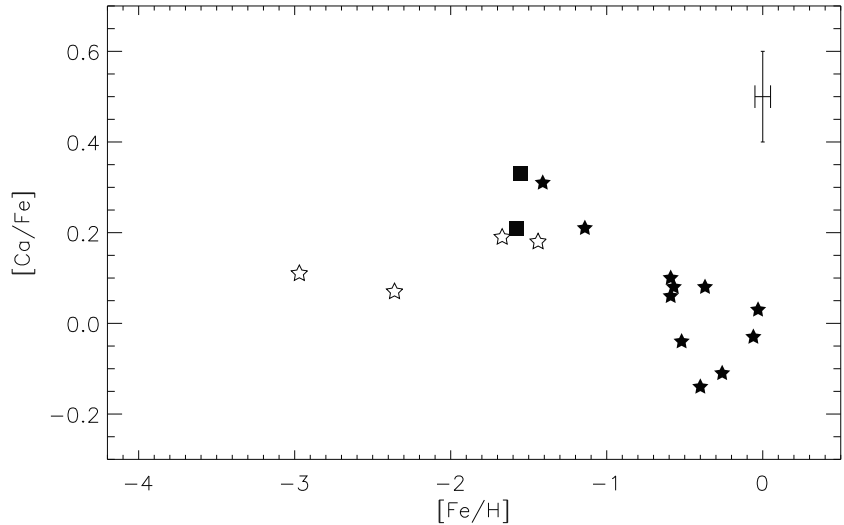}{6cm}{0.}{60.}{60.}{-175}{-240}
\vspace{0.7cm} \caption{Calcium to iron abundance ratio as a
function of metallicity for dSph stars: {\it solid stars} for Sgr
dSph field stars from Smecker-Hane, et al.  (1999) and the error
bar in the upper-right corner is an estimate of the average
$1-\sigma$ measurement error, {\it solid squares} for M54 stars in
the Sgr dSph from Brown, et al. (1999), {\it open squares} for
field stars in the Draco dSph from Shetrone, et al. (1998). }
\end{figure}

\nobreak  The merging history of the Galactic halo has been
constrained by comparing the colors of main-sequence turnoff
stars, metallicities, and the dark matter fractions of the Galaxy
versus the dSphs. However, the conclusions reached by Mateo (1996)
and Unavane, et al. (1996) do not agree. It is premature to apply
the element ratio test until we sample a range of dSph masses and
star-formation histories. However, we can say that not much of the
Galactic halo could be formed from the accretion of dSphs similar
to the Sgr dSph unless they merged many Gyr ago before the stars
reached [Fe/H] $\approx -0.6$, because the mean metallicity of the
Galactic halo is [Fe/H] $\approx -1.6$ and the 1--$\sigma$
dispersion is 0.65 dex. Recent studies have found Galactic halo
stars with [$\alpha$/Fe] $\approx +0.1$ (Nissen \& Schuster 1997,
Stephens 1999) similar to those found in the Draco dSph.
Fullbright (1999) is performing a large survey of halo stars that
will be ideal for estimating how much of the Galactic halo has
element ratios this low.

\begin{figure}
\plotfiddle{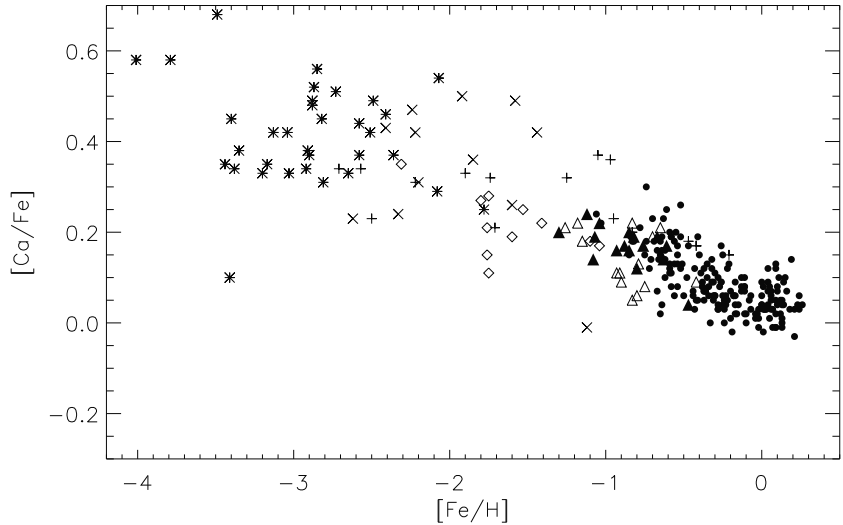}{6cm}{0.}{60.}{60.}{-175}{-240}
\vspace{0.7cm} \caption{Calcium to iron abundance ratio as a
function of metallicity for Galactic stars: {\it open circles} for
halo stars from McWilliam, et al.  (1995), {\it crosses} for
metal-poor stars from Magain (1987),  {\it pluses} for metal-poor
stars from Gratton \& Sneden (1991), {\it solid circles} for disk
stars from Edvardsson, et al.\ (1993), {\it solid triangles} for
disk stars and {\it open triangles} for halo stars from Nissen \&
Schuster (1997), {\it open diamonds} for stars with outer halo
kinematics from Stephens (1999).}
\end{figure}

The fact that the most metal-poor stars in the Sgr dSph have
[$\alpha$/Fe] nearly equal to the theoretical yield of Type~II SNe
and typical Galactic halo stars implies that the upper-mass end of
the initial mass function (IMF) in dSphs was similar to the
Galaxy's IMF. Measuring the element ratios for very metal-poor
stars in dSphs, [Fe/H] $\leq -2.0$, will be crucial for
determining the upper-mass end of the IMF and estimating the
energy available to power galactic winds. A study of the faint end
of the stellar luminosity function in the Ursa Minor dSph
(Feltzing, et al.\ 1999) concluded that the low-mass end of its
IMF was no different than that of the globular cluster M92. Thus
the IMF for dSphs may not be significantly different than the
Galactic IMF.

A significant fraction of dSph stars (2 of 11 field stars in the
Sgr dSph, 1 of 2 stars in M54, and 2 of 4 in Draco dSph) show
signs of having altering their primordial abundances of O, Na and
Al through proton burning. Sgr star \#1-73 shows this most
dramatically. Oxygen is depleted while Na and Al are enhanced.
This self-enrichment pattern was first identified in certain
Galactic globular clusters (see Kraft, et al.\ 1997, and
references therein), although it is not observed in any Galactic
field halo star (Shetrone, et al.\ 1996). The cause of the
difference is not yet understood, but we could use this as another
constraint on the merger history of the Galaxy.

By determining the ages and the chemical abundance ratios in dSphs
stars we also can gain new insights into stellar nucleosynthesis
by constraining the timescale on which an element is produced, and
thus the masses of stars which produce it. For example, we can
deduce the initial masses of the asymptotic-giant branch stars
that are the major sources of s-process elements. Also, we can
examine whether the trend in [Al/Fe] verses [Fe/H] for Galactic
field stars is driven by secondary production of Al or by a
metallicity-dependent yield.

\section {Conclusion}

Through detailed studies of the ages and chemical abundances of
the stellar population of dSph galaxies we will learn a great deal
about stellar nucleosynthesis, the physical processes that
regulate star formation, dwarf galaxy evolution, and the merging
history of our own Milky Way.

\vspace{2ex}

\acknowledgments TSH thanks the KO and CTIO staffs for their
excellent observing support, and the NSF for financial support
through grant AST-9619460. We also thank Andrew Carnegie and the
Keck Foundation for their generous gifts to public education and
astronomical research.

\nobreak

\end{document}